\documentclass[aps,twocolumn,prl,showpacs]{revtex4}
\usepackage{graphicx}
\usepackage{amssymb}
\usepackage{color}

\begin{document}

\title{Back and forth transfer and coherent coupling in a cold Rydberg dipole gas}

\author{Marcel Mudrich}
\email{mudrich@physik.uni-freiburg.de}
\author{Nassim Zahzam}
\author{Thibault Vogt}
\author{Daniel Comparat}
\author{Pierre Pillet}
\affiliation{Laboratoire Aim\'e Cotton, Campus d'Orsay B\^at. 505, 91405 Orsay, France}

\date{}
\begin{abstract}
Coupling by the resonant dipole-dipole energy transfer between cold cesium Rydberg atoms
is investigated using time-resolved narrow-band de-excitation spectroscopy. This technique
combines the advantage of efficient Rydberg excitation with high-resolution spectroscopy at
variable interaction times. 
Dipole-dipole interaction is observed spectroscopicaly as avoided level crossing. 
The coherent character of the process is linked to back and forth transfer in the 
$np+np\longleftrightarrow ns+(n+1)s$ reaction. 
Decoherence in the ensemble has two different origins: the atom motion induced by dipole-dipole
interaction and the migration of the $s$-Rydberg excitation in the environment of $p$-Rydberg atoms.
\end{abstract}


\maketitle

Cold ensembles of Rydberg atoms are particularly interesting quantum systems
because of the strong but controllable interactions between the atoms. 
Thermal motion of the atoms is mostly negligible on the timescale of Rydberg excitation.
In such "frozen Rydberg gases" resonant dipole-dipole energy transfer has been observed,
quite analogous to the migration of excitons in an amorphous solid~\cite{Mourachko,Anderson}. 
The many-body nature~\cite{MourachkoPRA} and coherent character~\cite{AndersonDephas}
of these resonant transfer processes have been clearly demonstrated.

This makes cold Rydberg gases promising candidate systems for fast quantum information schemes as
was suggested using tunable resonant dipole-dipole interaction associated with energy transfer resonances
~\cite{Lukin,Jaksch}. 
In particular, the Rydberg-Rydberg interaction may be 
exploited to induce a phase in a conditional phase-gate operation and to inhibit
excitation of pairs of Rydberg atoms within a mesoscopic volume. The 
signature of local blockade of Rydberg excitation due to long-range interaction
has been observed in experiments with narrow-band laser excitation~\cite{Tong,Singer} and line
broadening due to resonant dipole-dipole interactions has been observed using microwave
spectroscopy~\cite{Afrousheh}. It has recently been realized that dipole-dipole 
interactions between Rydberg atoms may also feature
dynamics with dramatic effects such as state redistribution phenomena, Penning ionization, and
subsequent plasma formation~\cite{Raithel,Robinson,LiPRA,Li}. 

The aim of this Letter is to understand the complete dynamics of reacting pairs of close Rydberg 
atoms in the environment of a dense Rydberg sample taking into account 
many-body processes~\cite{Mourachko,MourachkoPRA,Anderson} as well as
the relative motion of the interacting pairs of atoms~\cite{Fioretti, Li}. We show that
both effects lead to a loss of coherence in the evolution of the quantum Rydberg ensemble.
We consider the process
\begin{equation}
np_{3/2}+np_{3/2}\longleftrightarrow ns+(n+1)s
\label{eq:restransfer}
\end{equation}
with $n=24$ and $n=25$, which is resonant at electric fields $E_0=59.11$ and $E_0=44.03\,$V/cm, respectively.
Unless stated otherwise the magnetic component of the $np_{3/2}$-atoms is $\left|m_J\right|=1/2$.
In contrast to experiments employing narrow-band Rydberg excitation as
a means of probing interactions between the atoms 
\cite{Tong,Singer} we use broad-band pulsed laser excitation and time-delayed probing
by narrow-band de-excitation. This technique combines the advantage of high Rydberg
densities and high-resolution spectroscopy at variable interaction times. 

The Rydberg atoms are excited in a cloud of up to $10^7$ Cs atoms produced 
in a standard vapor-loaded MOT at residual gas pressure of $3\times 10^{-10}$\,mbar~\cite{Mourachko,Fioretti}.
At the trap position, a static electric field and a pulsed high voltage field can be applied by
means of a pair of electric field grids spaced by 15.7\,mm. The magnetic quadrupole field of the
MOT is switched off during the Rydberg excitation phase. For Rydberg excitation we use a pulsed
(7\,ns, $\sim 100\,\mu$J) dye laser with $\sim 10$\,GHz spectral width
running at 10\,Hz repetition rate. 
The Rydberg atoms are subsequently probed by a single-mode tunable Ti:Sa laser of up to 400\,mW output power.
The beams of the pulsed dye and Ti:Sa lasers are coaxially aligned and weakly focused into the
atomic cloud. 

\begin{figure}
\center
\includegraphics[width=7 cm]{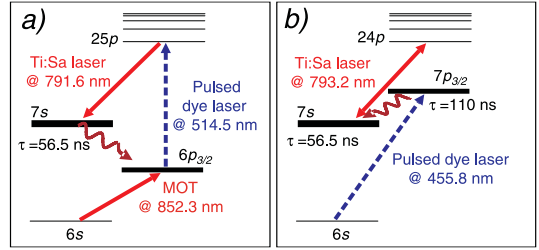}
\caption{(Color online) (a) Schematic representation of the scheme of pulsed excitation of $25p_{3/2}$ 
Rydberg states and cw de-excitation to the short-lived $7s$ state. (b) Combined pulsed and cw
excitation scheme for narrow-band excitation of $24p_{3/2}$ and subsequent de-excitation.} \label{fig:levels}
\end{figure}
The excitation scheme used in the first experiment presented in this Letter
is schematically depicted in Fig.~\ref{fig:levels} (a). Up to $6\times 10^5$ atoms are excited by the pulsed dye
laser from the 6$p_{3/2}$ upper MOT state to the 25$p_{3/2}$ Stark-Rydberg state. 
To fulfil the energy transfer resonance condition the electric field is tuned to a value 
such that the energy of a pair of $np$-atoms denoted by $\left| pp\right\rangle$ matches the energy of a pair of
$ns$- and $(n+1)s$-atoms $\left| ss'\right\rangle$ by the Stark effect.
In the two-atom picture resonant dipole-dipole interaction lifts degeneracy of the unperturbed states
which linearly combine to the coherent repulsive and attractive interaction states,
$\left|+\right\rangle =\left(\left| pp\right\rangle+\left| ss'\right\rangle\right)/\sqrt{2}$
and $\left|-\right\rangle =\left(\left| pp\right\rangle-\left| ss'\right\rangle\right)/\sqrt{2}$, respectively.
The corresponding energy curves $V_+(E)$ and $V_-(E)$
are the eigenvalues of the perturbed two-level Hamiltonian as a function of electric field $E$ 
\begin{equation}
\left(\begin{array}{cc}
V_{pp}(E) & \Delta V \\ \Delta V & V_{ss'}(E)
\end{array}\right),
\label{eq:ham}
\end{equation}
where $V_{pp}(E)$ and $V_{ss'}(E)$ stand for the energies of the unperturbed states $\left| pp\right\rangle$ and 
$\left| ss'\right\rangle$ and $\Delta V $ denotes the dipole-dipole interaction energy.
The coherent character of the transfer has been demonstrated using a Ramsey interference method
\cite{AndersonDephas}. In that work, the migration of the products of resonant energy transfer in 
the Rydberg ensemble is associated with observed dephasing of interference fringes.

In the experiment, up to 50\,\% of the population of $p$-atoms which are in the interacting magnetic sublevel 
are transformed into pairs of $s$ and $s'$-atoms within 300\,ns interaction time. Subsequently, the $25p$-Rydberg atoms
are coupled to the $7s$ state by the Ti:Sa laser. The Ti:Sa laser is applied continuously
during 1\,$\mu$s before selectively detecting the number of $25p$ and $26s$ atoms by applying 
a field ionizing high voltage pulse with $\sim 1\,\mu$s rise time and recording the ion signal
with gated integrators. Since the lifetime of the $7s$ state is short (56.5\,ns) compared with the Rydberg lifetime 
($\sim 10\,\mu$s), tuning the Ti:Sa laser into resonance leads to a drop of the number of 
detected Rydberg atoms. The Ti:Sa laser frequency is recorded using a commercial wavemeter
with 3\,MHz resolution (Angstr\"om WS-8).

\begin{figure}
\center
\includegraphics[width=9 cm]{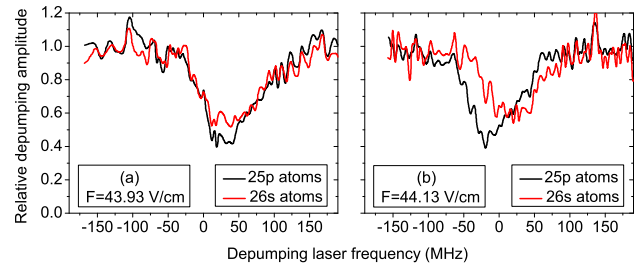}
\caption{(Color online) Typical depumping spectra of $25p_{3/2}$-atoms at two 
different values of the electric field $E=43.93$\,V/cm (a) and $E=44.13$\,V/cm (b). 
The frequency axis is shifted by 12632.0107\,cm$^{-1}$ which corresponds to the $25p$ line
position on energy transfer resonance ($F=44.03\,$V/cm). The $p$ and $s$-field
ionization signals are normalized to signals without depumping laser.} \label{fig:DepSpec}
\end{figure}

Typical depumping spectra of the $25p_{3/2}$ state at two 
different values of the electric field are displayed in Fig.~\ref{fig:DepSpec}.
The number of $25p$ and $26s$ Rydberg atoms is normalized to the number of excited atoms
without depumping laser, which is $6\times 10^5$ and $1.5\times 10^5$, respectively. 
The displayed spectra are the results of averaging over
4 individual scans and low-pass filtering to eliminate shot-to-shot fluctuations which
are in the range of 30\,\%. On resonance, the number of $25p$-atoms drops to about 40\,\%, which
is mainly due to the fact that both $\left| m_j\right| =1/2$ and $\left| m_j \right| =3/2$ components are
excited by the broad-band pulsed laser but only $\left| m_j\right|=1/2$ is coupled by the narrow-band depumping laser. 
Remarkably, we are also able to observe the reduction of $26s$ atoms by as much as 40\,\%, as
depicted in Fig.~\ref{fig:DepSpec}. This observation implies that resonant coupling of 
$s$-atoms back to $p$-atoms is also efficient.
However, the position of the depumping
resonance of the $s$-signal is shifted to higher frequencies when the electric field is 
tuned to values slightly higher than transfer resonance~(Fig.~\ref{fig:DepSpec}~(b)). Apparently, resonant coupling of
$s$-atoms back to $p$-atoms is efficient only when exciting the symmetric repulsive 
state $\left|+\right\rangle$, which in turn remains nearly perfectly coherent during the depumping period.
These results can be interpreted as a consequence of relative motion of the pairs of atoms
leading for the attractive case ($\left|-\right\rangle$) to $l$-mixing of states in the manifold~\cite{Raithel,Li,Fioretti}.
For pairs of atoms which are more closely spaced by a factor 5
than average the displacement within 1\,$\mu$s is about 10\,$\mu$m, which is comparable with the average distance between 
Rydberg atoms. This assumption is supported by recent work demonstrating the motion of atoms induced
by dipole forces~\cite{Li}. However, in our experiment we cannot directly observe such state mixing. 

\begin{figure}
\center
\includegraphics[width=9 cm]{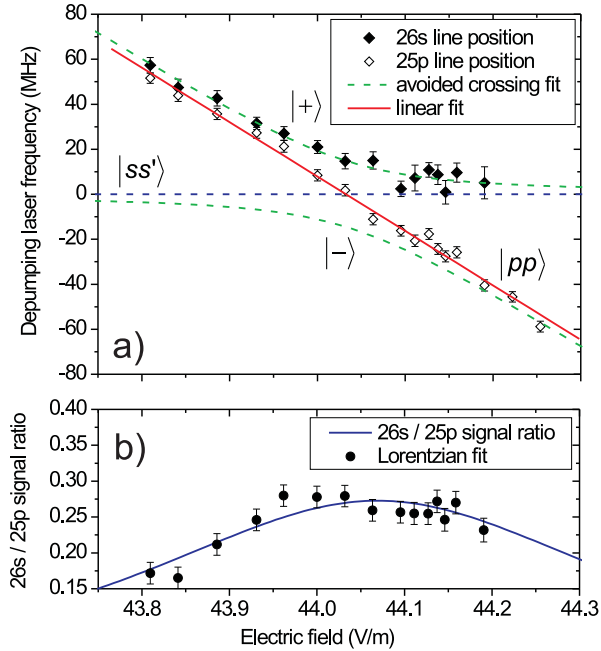}
\caption{(Color online) (a) Positions of the minima in the $p$ and $s$ spectra 
(open and filled diamonds, respectively) for various values
of the electric field as shown in Fig.~\ref{fig:DepSpec}. The straight line is a linear fit
of the positions of $p$-lines, the dashed lines indicate the interaction potential curves $V_+(E)$ and $V_-(E)$ from
Eq.~\ref{eq:ham} obtained
from a least-squares fit of $V_+(E)$ to the $s$-line positions. (b) The filled circles and the 
Lorentzian fit curve represent the number of $26s$ atoms relative to the number of $25p$ atoms 
without depumping laser.} \label{fig:AvoidCross}
\end{figure}

In Fig.~\ref{fig:AvoidCross} (a) the positions of $p$ and $s$-lines obtained from fitting Lorentzian functions
to the experimental spectra are plotted versus electric field. The corresponding $26s$-atom number normalized to the
$25p$-atom number without depumping laser is shown in Fig.~\ref{fig:AvoidCross} (b) including a Lorentzian fit curve.
The linear fit curve in Fig.~\ref{fig:AvoidCross} (a)  illustrates the Stark shift
of the $p$-state $V_{pp}$. The dashed lines result from fitting to the $s$-line positions the eigenvalue curve $V_+(E)$
of expression~(\ref{eq:ham}).
Free fit parameters are $V_{ss'}$, which is assumed to be independent of the electric field, and $\Delta V $.
The resulting value $\Delta V/h =15(3)\,$MHz has to be compared with the expected value, which can
be estimated by $\Delta V_{theo}\approx (\mu_{25s25p}\mu_{26s25p})/R^3$ in atomic units (a.\,u.),
where $\mu_{25s25p}=241$ a.\,u. and $\mu_{26s25p}=237$ a.\,u. denote the transition dipole moments, 
and $R$ stands for the mean distance between interacting Rydberg atoms.
The resulting value $\Delta V_{theo}/h\approx 0.5\,$MHz
is much smaller than the experimental one which is attributed to many-body effects 
inducing a wide energy band as in a disordered solid 
rather than a splitting of degenerate levels in the simple two-atom picture~\cite{Anderson, AndersonDephas, Mourachko, MourachkoPRA}.
We have carefully checked that the effect was not an artefact due to the shape of the field ionization pulse
which could lead to (a-)diabatic following of the Stark manifold.

\begin{figure}
\center
\includegraphics[width=8 cm]{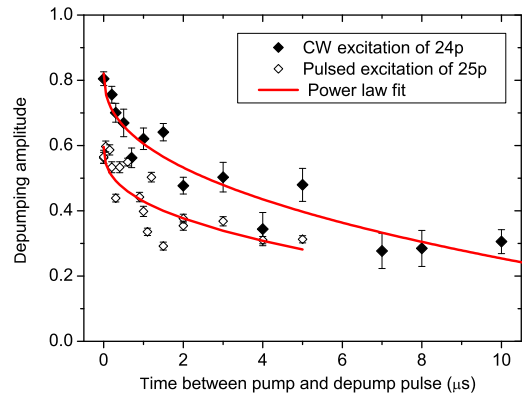}
\caption{(Color online) Temporal evolution of the amplitude of depletion of $25s$ and $26s$-populations
(filled and open diamonds, respectively).}
\label{fig:deltat}
\end{figure}

Coherence of the dipole coupled interaction state $\left| +\right\rangle$ is interrogated by 
measuring the back and forth coupling efficiency for variable evolution times of the Rydberg sample. For this,
the narrow-band depumping pulse is time-delayed with respect to the excitation pulse. The field ionizing high-voltage ramp
is applied at constant time 10\,$\mu$s after the excitation pulse.
The open diamonds in Fig.~\ref{fig:deltat} represent the depletion amplitudes of $26s$-atoms
which are obtained from Lorenzian fits of the $26s$-lines. The plotted amplitudes are normalized to the $26s$-signal 
recorded without depumping pulse.  The initial depumping amplitude of up to 60\,\% continuously decreases down to roughly
30\,\% at 1\,$\mu$s and longer interaction time. 
We interpret this behavior as a loss of the coherence after $~1\,\mu$s, due to the many-body migration of the 
products $s$, $s'$ of the reaction (\ref{eq:restransfer}) through the inhomogeneous $p$-Rydberg ensemble~\cite{AndersonDephas}.
For longer times, the ensemble is in an incoherent sum of $p$ and $s$, $s'$-populations, where secondary reactions 
between $s$ and $s'$ atoms explain the residual depumping. The contribution
to decoherence from the thermal motion of the Rydberg atoms is expected to be negligible since the average displacement
of an atom in $1\,\mu$s is only $\sim 0.1\,\mu$m. 

In a complementary experiment, the dynamics of resonant excitation transfer is investigated on the timescale of several 
microseconds. Rydberg atoms are now excited by a combined scheme of pulsed excitation, spontaneous emission, and
cw-excitation of the Rydberg levels as illustrated in Fig.~\ref{fig:levels} (b). The branching ratio for transitions
from $7p_{3/2}$ to $7s$ intermediate state is 44\,\%~\cite{kurucz}. Atoms are excited selectively to 
the $24p_{3/2},\left| m_j\right| =1/2$ Rydberg
state by switching on the Ti:Sa laser during $t_1 =0.3\,\mu$s after the dye laser pulse by means of an acousto-optic
modulator. This pump pulse transfers the maximum number of atoms into the Rydberg
state. After an off-period of $\Delta t = 0.1$-$10\,\mu$s a second pulse of $t_2 =1\,\mu$s duration is applied 
for down-stimulating the Rydberg atoms to the short-lived $7s$-state as in the previous experiment. The field
ionizing high-voltage pulse is applied at a fixed time of 5\,$\mu$s after the dye laser pulse for the measurements with
$\Delta t=0$-$3\,\mu$s and 15\,$\mu$s after the dye laser pulse for the measurements with
$\Delta t=4$-$10\,\mu$s.

\begin{figure}
\center
\includegraphics[width=8 cm]{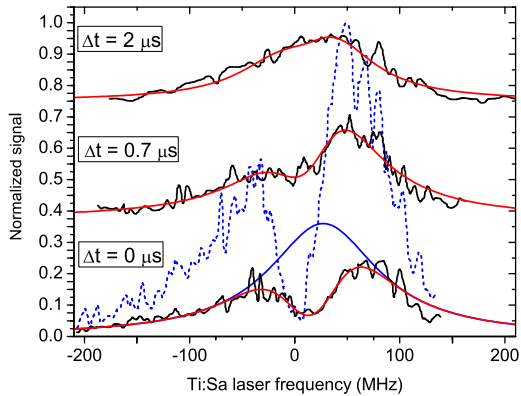}
\caption{(Color online) Typical $24p$ and $25s$-spectra obtained after excitation and time-delayed de-excitation 
of $24p$ Rydberg atoms for various delay times $\Delta t$. The frequency axis is shifted by
12608.0256\,cm$^{-1}$ which corresponds to the position of maximum de-excitation of $24p$-atoms.
The smooth lines result from fitting the spectra with 
a heuristic model function (see text).}
\label{fig:24pspec}
\end{figure}

The observed $24p$-resonance is depicted as a dashed line in Fig.~\ref{fig:24pspec} 
for an electric field $E=59.11\,$V/cm which corresponds to maximum excitation transfer. 
Depumping of the $p$-atoms takes place at the center
of the excitation line for two reasons. On the one hand, the Rabi frequency is largest on resonance which leads to
maximum depletion of the Rydberg. 
On the other hand, a small number of photo-ions ($\lesssim 100$) is created by the Ti:Sa laser, 
presumably by photo-ionization of the $7p_{3/2}$-state. The space charge from these ions induces inhomogeneous Stark
broadening during Rydberg excitation. 
However, the ions are accelerated out of the Rydberg sample within 300\,ns at 59.11\,V/cm and the depumping line
is unaffected by photo-ions. The depumping minimum nearly reaches zero in this scheme since all excited Rydberg
atoms are subject to subsequent de-excitation.

The $s$-lines for three different off-periods of the Ti:Sa laser are plotted vertically shifted from each other for 
the sake of distinctiveness. The vertical scale indicates the $s-$signal amplitude relative to the maximum $p-$signal.
Up to 50$\,\%$ of $p$-atoms are converted into $s$-atoms, which requires many-body interactions to govern the excitation
transfer process~\cite{Anderson,Mourachko,MourachkoPRA}.
When comparing the depumping minima of $p$ and $s$-lines one notices that again depletion of the $s$-atoms is efficient
only on the high-frequency side of the spectral line. In fact, the $p$ and $s$-depumping line positions 
qualitatively reproduce a similar $E$-field dependence as the one shown in Fig.~\ref{fig:AvoidCross}. 
However, the fitted interaction energy amounts to
$\Delta E\approx 1\,$MHz which can be explained by a factor 6 lower Rydberg density than in the 
first experiment using pulsed Rydberg excitation. 

For an increasing off-period $\Delta t$
of the Ti:Sa laser the depumping efficiency clearly deminishes. In fact we observe a hole in the spectrum
only up to $\Delta t\sim 1.5\,\mu$s, and a reduced signal amplitude for longer times. 
In order to quantify the depumping dynamics the $s$-lines are fitted by the heuristic model function
$N_s(\nu)=L(\nu )\times \left(1-G(\nu )\right)$,
where $L$ and $G$ stand for Lorentzian and Gaussian functions. During the fitting procedure the peak positions 
are held fixed as well as the Lorentzian peak amplitude which is obtained from fitting the $s$ line without
depumping laser as indicated by the smooth curve of Lorentzian shape.
The resulting Gaussian amplitude factor which may vary between zero (no depumping) and
unity (depumping down to zero) is plotted versus $\Delta t$ as filled diamonds in Fig.~\ref{fig:deltat}.
Depumping efficiency
drops fastest during the first microsecond which reflects loss of coherence of the system as previously mentioned. 
In the entire time range up to 10\,$\mu$s, the data are modelled by a power-law decay which is plotted as smooth line 
to guide the eye. When comparing the data for broad-band and narrow-band Rydberg excitation one notices different
initial depumping amplitudes which has already been discussed, but time evolution qualitatively agrees
for both experiments.

In conclusion, we have reported on narrow-band time-resolved spectroscopy of a gas of interacting Rydberg atoms,
leading to a better understanding of decoherence of an interacting Rydberg ensemble. 
Excitation transfer at the attractive potential branch is not observed,
presumably due to state-mixing dipole-induced collisions.  
This first result indicates the frontier between a frozen Rydberg gas where the thermal motion of atoms is negligible
and a dipole gas where the motion between atoms induced by the dipole forces dominates the dynamics.
In such a dipole gas, any initial coherence is rapidly destroyed. Decoherence of the repulsive interaction state takes
place on a time scale of $\sim 1\,\mu$s which can be attributed to the far migration of the products of reacting pairs
of atoms. This second result seems to confirm the dynamics of coherence of such quantum Rydberg ensembles as observed
by~\cite{AndersonDephas}. 
Further experiments and new theoretical developments are necessary to reach complete understanding of such quantum systems.

This work is in the frame of the European Research and Training
Network COLMOL (HPRN-CT-2002-00309) and QUACS (HPRN-CT-2002-00309). One of the
authors (M.~M.) is supported by COLMOL. The autors acknowledge 
fruitful discussions with T.~Gallagher, V.~Akulin and E.~Brion.

\end{document}